\documentstyle[sprocl,epsfig,axodraw,color]{article}

\input{paperdef}
\begin{document}

\thispagestyle{empty}
\setcounter{page}{0}
\def\thefootnote{\fnsymbol{footnote}}

\begin{flushright}
CERN--PH--TH/2004--159\\
hep-ph/0408269 \\
\end{flushright}

\vspace{1cm}

\begin{center}

{\large\sc {\bf Loop Calculations: Summary}}
\footnote{plenary talk given at the ``International Conference on
  Linear Colliders'',
April 2004, \mbox{}~~~~~~Paris, France}

\vspace{1cm}

{\sc S.~Heinemeyer$^{\,}$%
\footnote{
email: Sven.Heinemeyer@cern.ch
}%
}

\vspace*{0.5cm}

CERN TH Division, Dept.\ of Physics, CH~1211 Geneva 23, Switzerland

\end{center}

\vspace*{1cm}

\begin{abstract}
Current and future colliders will provide high precision experimental
data. In order to use the high experimental precision it has to be
matched with theoretical predictions at the same level of accuracy or
better. This involves the calculation of loop corrections at
increasingly higher order. We briefly 
review the status of the field, especially in view of the calculations
presented at this conference. We give an outlook about the
theoretical requirements for the anticipated precision of a future 
$e^+e^-$ linear collider.
\end{abstract}

\def\thefootnote{\arabic{footnote}}
\setcounter{footnote}{0}

\newpage


\title{LOOP CORRECTIONS: SUMMARY}

\author{ SVEN HEINEMEYER }

\address{CERN TH Division, Dept.\ of Physics, CH~1211 Geneva 23, Switzerland}


\maketitle\abstracts{
Current and future colliders will provide high precision experimental
data. In order to use the high experimental precision it has to be
matched with theoretical predictions at the same level of accuracy or
better. This involves the calculation of loop corrections at
increasingly higher order. We briefly 
review the status of the field, especially in view of the calculations
presented at this conference. We give an outlook about the
theoretical requirements for the anticipated precision of a future 
$e^+e^-$ linear collider.
}


\section{Introduction}
\label{sec:intro}

Past experiments have reached a precision that their results disagreed
with the lowest-order Standard Model (SM) calculation. One example is
the mass of the $W$~boson, which can be predicted (at tree-level) in
terms of the $Z$~boson mass, the Fermi constant and the fine structure
constant. However, only by taking the full
1-loop and leading 2- (and 3-) loop corrections into account
(see e.g.\ \citeres{deltarSM2l,delrhoSM3l} for recent calculations),
the experimentally measured value from LEP2~\cite{LEP2} is in
agreement with the SM prediction. Another example is the cross section for 
$e^+e^- \to W^+W^-$. Only by the inclusion of the double-pole
approximation 1-loop result~\cite{eeWWSM1l} the experimental result
from LEP2~\cite{LEP2} and SM prediction are in good agreement.
As a last example we take the anomalous magnetic moment of the
muon. The final experimental result of the Brookhaven 
``Muon $g-2$ Experiment'' (E821)~\cite{g-2exp} has a precision of 
$6 \times 10^{-10}$, the same size as (and thus being 
sensitive to) the SM electroweak (EW) 2-loop 
contributions~\cite{g-2SM2lEW}.

The currently operating Tevatron and other (low energy) experiments
have the potential to improve current precisions. The upcoming LHC,
besides being a discovery machine for new physics, will also improve
precision measurements. Finally the prospective $e^+e^-$ linear
collider (LC) will provide measurements of masses,
couplings and cross sections at (or even below) the per-cent
level~\cite{LCexp}. A special example is the GigaZ option, which will
determine the $W$~boson mass with an error of $7 \mev$ and the
effective leptonic weak mixing angle with a precision of 
$1.3 \times 10^{-5}$, see \citeres{GigaZexp,blueband} and refs.\
therein. These anticipated future  
high precision measurements can only be utilized if they are matched
with a theory prediction at the same level of accuracy or better.
These theory predictions have to be obtained in the model under
investigation (e.g.\ the SM, or the Minimal Supersymmetric SM (MSSM),
for recent overviews see \citeres{SDAmsterdam,MSSMtheo}).


\section{Status Of The Field}

\definecolor{Blue}{named}{Blue}
\definecolor{Red}{named}{Red}
\definecolor{Green}{named}{PineGreen}
\definecolor{Black}{named}{Black}
\definecolor{Magenta}{named}{Magenta}
\definecolor{Royal}{named}{RoyalBlue}
\definecolor{Orange}{named}{Orange}
\definecolor{Apricot}{named}{Apricot}
\definecolor{Purple}{named}{Purple}
\newcommand{\black}[1]{\color{Black}#1 \color{Black}}
\newcommand{\gray}[1]{\color{Gray}#1 \color{Black}}
\newcommand{\red}[1]{\color{Red}#1 \color{Black}}
\newcommand{\blue}[1]{\color{Blue}#1 \color{Black}}
\newcommand{\green}[1]{\color{Green}#1 \color{Black}}
\newcommand{\magenta}[1]{\color{Magenta}#1 \color{Black}}
\newcommand{\royal}[1]{\color{Royal}#1 \color{Black}}
\newcommand{\orange}[1]{\color{Orange}#1 \color{Black}}
\newcommand{\apricot}[1]{\color{Apricot}#1 \color{Black}}
\newcommand{\purple}[1]{\color{Purple}#1 \color{Black}}

\begin{figure}[htb!]
\vspace{-.5cm}
\begin{center}
\setlength{\unitlength}{0.5pt}
\begin{picture}(650,400)
  \LinAxis(0,0)(312.5,0)(10.999,1,-1,0,1)
  \LongArrow(0,0)(312.5,0)
  \LinAxis(0,0)(0,180)(3.999,1,+1,0,1)
  \LongArrow(0,0)(0,180)
  \Text(-10,  0)[c]{0}
  \Text(-10,090)[c]{1}
  \Text(-10,180)[c]{2}
  \Text(-10,270)[c]{3}
  \Text(20,380)[c]{\# loops}
  \Text(  0,-20)[c]{0}
  \Text( 57,-20)[c]{1}
  \Text(114,-20)[c]{2}
  \Text(171,-20)[c]{3}
  \Text(228,-20)[c]{4}
  \Text(285,-20)[c]{5}
  \Text(342,-20)[c]{6}
  \Text(399,-20)[c]{7}
  \Text(456,-20)[c]{8}
  \Text(513,-20)[c]{9}
  \Text(570,-20)[c]{10}
  \Text(630,-17)[c]{\# legs}
  \CBox(  1,  1)( 59, 95){Green}{Green}
  \CBox( 59,  1)(117, 50){Green}{Green}
  \CBox(117,  1)(231,  5){Green}{Green}
\CBox(150,182)(170,187){Green}{Green}
  \Text(350,370)[l]{\green{Technique well established}}
\Text( 20,-50)[c]{\scriptsize vacuum diag.}
\Text( 10,-70)[c]{\scriptsize $\De\rho$}
\Text(120,-50)[c]{\scriptsize self-energies}
\Text(120,-70)[c]{\scriptsize $\De r$, masses}
\Text(179,-90)[c]{\scriptsize $1 \to 2$, $\sweff$}
\Text(234,-50)[c]{\scriptsize $2 \to 2,\, 1 \to 3$}
\Text(234,-70)[c]{\scriptsize Bhabha}
\Text(291,-90)[c]{\scriptsize $2 \to 3$}
\Text(348,-50)[c]{\scriptsize $ee \to 4f$}
\Text(406,-70)[c]{\scriptsize $ee \to 4f + \ga$}
\Text(462,-50)[c]{\scriptsize $ee \to 6f$}
\CBox(  1, 95)( 59,142){Blue}{Blue}
\CBox(  1,142)(  3,175){Blue}{Blue}
\CBox( 59, 50)(117, 95){Blue}{Blue}
\CBox(117,  5)(144, 50){Blue}{Blue}
\CBox(231,  1)(290,  5){Blue}{Blue}
\CBox(150,164)(170,169){Blue}{Blue}
\Text(350,330)[l]{\blue{Partial results/special cases}}
\CBox(  0,120)( 15,135){Red}{Red}
\CBox(  2,122)( 13,133){Red}{White}
\Text( 18,255)[c]{\red{1}}
\CBox(100, 78)(115, 93){Red}{Red}
\CBox(102, 80)(113, 91){Red}{White}
\Text(219,170)[c]{\red{2}}
\CBox(127, 33)(142, 48){Red}{Red}
\CBox(129, 35)(140, 46){Red}{White}
\Text(273, 80)[c]{\red{3}}
\CBox(155,140)(170,155){Red}{Red}
\CBox(157,142)(168,153){Red}{White}
\Text(330,295)[c]{\red{\#}}
\Text(350,290)[l]{\red{Some results from the front}}
\end{picture}
\end{center}
\vspace{1.5cm}
\caption{
The current status of the field of loop calculations is shown. The
medium shaded (green) area shows the well established techniques. The dark
shaded (blue) area corresponds to the number of legs and loops where
partial results or calculations for special cases are known. Also
indicated with squares and numbers are some recently obtained results.
}
\label{fig:status}
\end{figure}
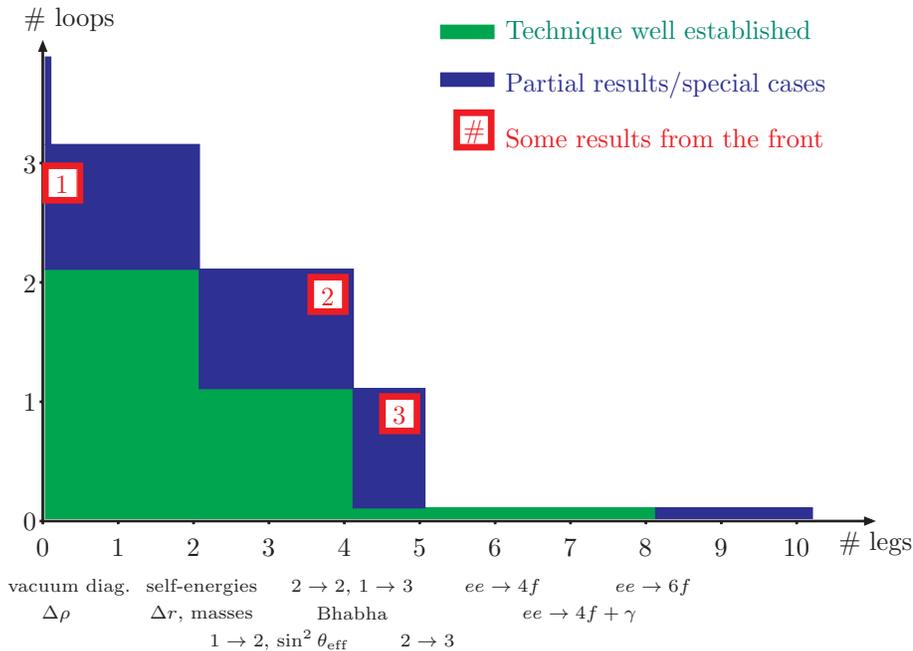

The status of the field of loop calculations is briefly summarized in
\reffi{fig:status}.
%
%
The complication of a higher-order loop calculation increases with
the number of loops as well as with the number of external legs. On
the other hand, it also increases with the number of (mass) scales
appearing in
the loop integral. While one-scale integrals (as occur e.g.\ in QCD)
are usually the easiest possibility for a certain loop topology, two
or more scales make the evaluation increasingly difficult. This poses
a special problem in EW calculations, where many independent mass
scales can appear in a single loop diagram. 
In \reffi{fig:status} on the horizontal axis the
number of legs, and on the vertical one the number of loops is
shown. Accordingly, the number of scales has to be kept in mind for
each individual case presented below.

The medium shaded (green) area in \reffi{fig:status} displays the
number of loops and legs for which the techniques are meanwhile well
established, even for an arbitrary number of scales. 
For these cases often public algebraic computer codes exist that
do the main part of the calculation itself, for an overview see
\citere{SDAmsterdam}.
The dark shaded (blue) area corresponds to the number of loops and legs
for which partial results or calculations for special cases have been
performed. This represents today's frontier of the field of loop
calculations. For the sake of brevity we mention only three results from
the front, indicated by the squares and numbers. 
\#~1~is an example for a 3-loop calculation of vacuum diagrams (i.e.\
with no external legs) and one scale: the leading 3-loop EW
corrections to the $\rho$~parameter in the
SM~\cite{delrhoSM3l}. Translated to precision observables, these
corrections eliminated a theoretical uncertainty of $4-5 \mev$ in the
SM prediction of $\MW$ and $2-3 \times 10^{-5}$ in the effective weak
leptonic mixing angle, $\sweff$, see \refse{sec:intro}.
\#~2~indicates the progress made over the recent years in the
evaluation of (massless) 2-loop box calculations obtained by several
groups~\cite{2lbox}. These calculations are especially important e.g.\
for jet physics at the LHC and LC.
\#~3~represents the progress made in the last two
years in the evaluation of the full 1-loop EW corrections
(implying an arbitrary number of scales) to 
$2 \to 3$ processes. The full 1-loop calculation of 
$e^+e^- \to t \bar t H$ has been obtained by three independent
groups~\cite{eettH}. These corrections are indispensable in order to
match the anticipated precision of a future LC.
Finally we would like to mention the progress made in the automated
reduction of loop integrals to basic sets of integrals. The
Integration By Parts (IBP) method~\cite{ibp} allows to obtain a set of
linear equations for different loop integrals. Following the method
proposed in \citere{laporta}, recently the first public code became
available~\cite{air} that (in principle) can perform the reduction to
master integrals for arbitrary loop diagrams.


\section{Contributions In Paris}

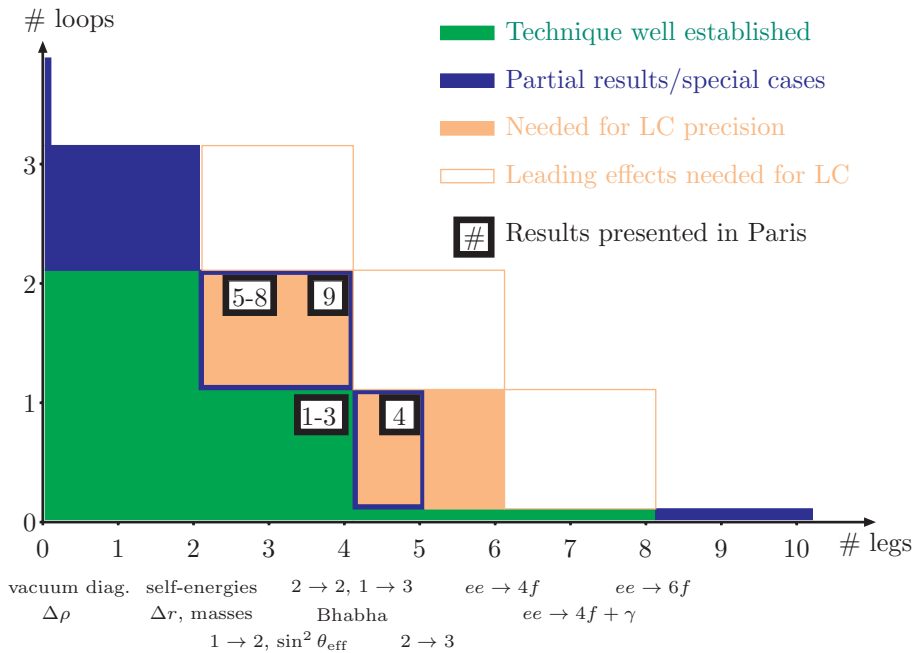
\begin{figure}[htb!]
\vspace{-.5cm}
\begin{center}
\setlength{\unitlength}{0.5pt}
\begin{picture}(650,400)
  \LinAxis(0,0)(312.5,0)(10.999,1,-1,0,1)
  \LongArrow(0,0)(312.5,0)
  \LinAxis(0,0)(0,180)(3.999,1,+1,0,1)
  \LongArrow(0,0)(0,180)
  \Text(-10,  0)[c]{0}
  \Text(-10,090)[c]{1}
  \Text(-10,180)[c]{2}
  \Text(-10,270)[c]{3}
  \Text(20,380)[c]{\# loops}
  \Text(  0,-20)[c]{0}
  \Text( 57,-20)[c]{1}
  \Text(114,-20)[c]{2}
  \Text(171,-20)[c]{3}
  \Text(228,-20)[c]{4}
  \Text(285,-20)[c]{5}
  \Text(342,-20)[c]{6}
  \Text(399,-20)[c]{7}
  \Text(456,-20)[c]{8}
  \Text(513,-20)[c]{9}
  \Text(570,-20)[c]{10}
  \Text(630,-17)[c]{\# legs}
\Text( 20,-50)[c]{\scriptsize vacuum diag.}
\Text( 10,-70)[c]{\scriptsize $\De\rho$}
\Text(120,-50)[c]{\scriptsize self-energies}
\Text(120,-70)[c]{\scriptsize $\De r$, masses}
\Text(179,-90)[c]{\scriptsize $1 \to 2$, $\sweff$}
\Text(234,-50)[c]{\scriptsize $2 \to 2,\, 1 \to 3$}
\Text(234,-70)[c]{\scriptsize Bhabha}
\Text(291,-90)[c]{\scriptsize $2 \to 3$}
\Text(348,-50)[c]{\scriptsize $ee \to 4f$}
\Text(406,-70)[c]{\scriptsize $ee \to 4f + \ga$}
\Text(462,-50)[c]{\scriptsize $ee \to 6f$}

\CBox(  1,  1)( 59, 95){Green}{Green}
\CBox( 59,  1)(117, 50){Green}{Green}
\CBox(117,  1)(231,  5){Green}{Green}
\CBox(  1, 95)( 59,142){Blue}{Blue}
\CBox(  1,142)(  3,175){Blue}{Blue}
\CBox( 59, 50)(117, 95){Blue}{Blue}
\CBox(117,  5)(144, 50){Blue}{Blue}
\CBox(231,  1)(290,  5){Blue}{Blue}
\CBox( 61, 52)(115, 93){Apricot}{Apricot}
\CBox(119,  7)(142, 48){Apricot}{Apricot}
\CBox(144,  5)(174, 50){Apricot}{Apricot}
\CBox( 60, 95)(117,142){Apricot}{White}
\CBox(117, 50)(174, 95){Apricot}{White}
\CBox(174,  5)(231, 50){Apricot}{White}

\CBox(150,182)(170,187){Green}{Green}
\Text(350,370)[l]{\green{Technique well established}}
\CBox(150,164)(170,169){Blue}{Blue}
\Text(350,334)[l]{\blue{Partial results/special cases}}
\CBox(150,146)(170,151){Apricot}{Apricot}
\Text(350,298)[l]{\apricot{Needed for LC precision}}
\CBox(150,128)(170,133){Apricot}{White}
\Text(350,262)[l]{\apricot{Leading effects needed for LC}}
\CBox(155,100)(170,115){Black}{Black}
\CBox(157,102)(168,113){Black}{White}
\Text(330,215)[c]{\black{\#}}
\Text(350,218)[l]{\black{Results presented in Paris}}

\CBox( 95, 33)(115, 48){Black}{Black}
\CBox( 97, 35)(113, 46){Black}{White}
\Text(212, 80)[c]{\black{1-3}}
\CBox(127, 33)(142, 48){Black}{Black}
\CBox(129, 35)(140, 46){Black}{White}
\Text(273, 80)[c]{\black{4}}
\CBox( 68, 78)( 88, 93){Black}{Black}
\CBox( 70, 80)( 86, 91){Black}{White}
\Text(160,170)[c]{\black{5-8}}
\CBox(100, 78)(115, 93){Black}{Black}
\CBox(102, 80)(113, 91){Black}{White}
\Text(220,171)[c]{\black{9}}

\end{picture}
\end{center}
\vspace{1.5cm}
\caption{
Besides the current status of the field and partial/special results
(see also \reffi{fig:status}), the light shaded (orange) area shows
what will be needed to match the anticipated LC precision.
The squares and numbers indicate the contributions presented at this
conference. 
}
\label{fig:paris}
\end{figure}

In order to match the anticipated experimental precision of a future
LC, the field of loop calculations has to advance substantially, as
will be further discussed in the next section. The necessary
improvement is indicated in \reffi{fig:paris} as the light shaded
(orange) areas, which will have to be under full control for the LC
precision. Some advance has been presented at this conference, which
is shown as black rectangles (and numbers). 

Three calculations of 1-loop corrections to $2 \to 2$ processes have
been presented. 
A SM (re)calculation of deep inelastic neutrino
scattering as observed in the NuTeV experiment~\cite{nutev} has been
performed~\cite{arbuzov}, using the automized SAN-C
system~\cite{sanc}. The comparison with another recent
cacluation~\cite{SDnutev} is still ongoing.
For the same process the complete calculation of all SUSY 1-loop
contributions has been performed~\cite{brein}, finding that SUSY cannot
explain the NuTeV anomaly, see also \citere{nutevSUSY}. 
Moreover, a new recursive algorithm for the
numerical evaluation of 1-loop tensor integrals has been
presented~\cite{pittau}. This algorithm is supposed to be applicable
to 1-loop diagrams with an arbitrary number of external legs, but 
so far has been applied to four.

A recalculation of the 1-loop QED corrections to $e^+e^- \to 2f + \ga$
(i.e.\ with five external legs, however, only box and triangle
diagrams contribute) in the 
SM has been presented~\cite{yost}. This process contributes to Bhabha
scattering, which as to be under 
control, since it serves as a luminosity monitor at the LC. 
The comparison with different other results showed agreement up to
$10^{-5}$. 

Four new calculations with three external legs at the 2-loop level
have been presented: 
First, the evaluation of EW logs of the type
$\al^L \log^K (s/M^2)$ for $1 \to 2$ SM processes has been
automized~\cite{pozzorini} (where $s$ is the center-of-mass energy,
$M$ is the mass of a SM gauge boson, and $s \gg M^2$). 
Second, an example for the evaluation of the (subleading) Sudakov logs
in EW processes has been performed~\cite{kuehn}. The form
factor for a $1 \to 2$ process within a massive $U(1)$ theory has been
evaluated in the high-energy limit, being of the form
$\al^n \, \sum_{k = 0}^4 \log^{2n - k} (s/M^2)$ for $n = 2$. 
Third, the evaluation of all 2-loop diagrams with a closed fermion
loop within the SM for $\sweff$ has been
presented~\cite{awramik}. This calculation is especially relevant,
since $\sweff$ is an important precision observable for the indirect
determination of the Higgs boson mass~\cite{LEP2,blueband}. The new
calculation eliminates a theoretical uncertainty of 
$\sim 4 \times 10^{-5}$ (see \refse{sec:intro}). 
Fourth, new SUSY 2-loop corrections to the anomalous magnetic moment of
the muon have been shown~\cite{doink}. Since the SM shows a 2-3$\,\si$
deviation from the experimental result, SUSY is a good candidate to
accommodate this ``discrepancy''. The 2-loop corrections are necessary
to achieve a precision of the MSSM prediction of $1 \times 10^{-10}$,
as has been done for the corresponding SM evaluation. It has been
shown that the new evaluation can shift $2\si$ exclusion bounds
in the MSSM parameter space substantially. 

Finally the progress in the evaluation of 2-loop QED corrections for Bhabha
scattering has been presented~\cite{gluza}, being important for a
future luminosity monitor at the LC, see above. This new calculation
with one or two scales in each integral requires the evaluation of
more than 40 new 2-loop box master integrals and has not been
completed yet. This calculation is also an example of the application
of the methods of \citeres{ibp,laporta}.


\section{Conclusions And Outlook}

The requirement of loop calculations to match the LC precision is
shown in \reffi{fig:paris} as the light shaded (orange) area. Light
(orange) boxes indicate for which number of loops and legs at least
leading effects will have to be available. One can see that the Paris
contributions certainly advanced the field towards the required
precision. However, they constitute the evaluation of special
cases (sometimes possible due to a fortunate kinematical situation).

A lot of work remains to be done.
Currently we are far away from e.g.\ a full EW 
evaluation of $2 \to 2$ processes at the 2-loop level (though
considerable progress in the corresponding QED and QCD processes has
been made). 
On the other hand, the high experimental precision achievable at a LC
will be worthless if it cannot be matched with theoretical precisions
at the same level of accuracy. Therefore the high-energy physics
community has to continously support the advance in this field to have
the calculations at hand when the LC starts. Theoretical calculations
should be viewed as an essential part of all current and future high-energy
physics programs.


\section*{Acknowledgements}

We thank S.~Dittmaier for helpful discussions.


\end{document}